# A Hybrid Method for Low-Resource Named Entity Recognition

Do Minh Duc[1], Quan Xuan Truong[2], Viet Tran Hong[3], Le Hoang Anh[4], Mac Thi Minh Tra[5], Nguyen Van Thuy[6], Le Hai Ha[7], Vinh Nguyen Van[8,*]

[1,2,3,8] *Vietnam National University, Hanoi*

[4,5,6] *Center for Biodiversity Monitoring and Investigation*

[7] *Hanoi University of Science and Technology*

**Abstract**

Named Entity Recognition (NER) is a critical component of Natural Language Processing with diverse applications in information extraction and conversational AI. However, NER in specific domains for low-resource languages faces challenges such as limited annotated data and heterogeneous label sets. This study addresses these issues by proposing a hybrid neurosymbolic framework that integrates rule-based processing with deep learning models for Vietnamese NER. The core idea involves a two-stage pipeline: first, a rule-based component reduces label complexity by grouping relational and special categories; second, pre-trained language models are fine-tuned for high-precision extraction. A post-processing module is then utilized to restore fine-grained labels, preserving expressiveness for application-level usability. To mitigate data scarcity, a scalable data augmentation strategy leveraging Large Language Models (LLMs) is introduced to expand the label set without full re-annotation—a significant novelty of this work. The effectiveness of this method was evaluated across five specific-domain datasets, including logistics, wildlife, and healthcare. Experimental results demonstrate substantial improvements over strong RoBERTa-based baselines. Specifically, the proposed system achieved F1 scores of 90% in Customer Service (up from 83%), 84% in GAM (up from 73%), 83% in AI Fluent (up from 80%), 94% in PhoNER_Covid19 (up from 91%), and 60% in Rare Wildlife (up from 36%). These findings confirm that the hybrid approach effectively captures the linguistic complexity of Vietnamese and contextual nuances in specialized domains, offering a robust contribution to low-resource NER research.

## 1. Introduction

Named Entity Recognition (NER) is a fundamental task in Natural Language Processing (NLP) [1], defined as the process of identifying and classifying named entities within unstructured text into predefined categories such as person names, organizations, locations, and miscellaneous entities [2]. This task plays a pivotal role in numerous NLP systems by transforming unstructured text into structured information, thereby enhancing the capabilities of applications like information retrieval, chatbots, and question answering systems. In recent years, the advent of deep learning has significantly advanced the performance of NER systems, with models such as BERT and its variants achieving state-of-the-art results across multiple benchmarks [3]. These advanced models use large amounts of data to learn the context of words, allowing them to perform better than traditional methods [4]. Moreover, multilingual and domain-specific adaptations of these models have further extended the applicability of NER in diverse settings such as biomedical, legal, and e-commerce domains [5].

In the context of Vietnamese, however, NER presents additional challenges due to language-specific characteristics such as the lack of explicit word boundaries, frequent use of compound words, and flexible syntactic structures [6]. These features complicate both word segmentation and contextual disambiguation, which are crucial for accurate entity recognition. Early approaches to Vietnamese NER primarily relied on rule-based methods and statistical models like Conditional Random Fields (CRF) [7]. The introduction of deep learning techniques, particularly Bidirectional Long

*Corresponding author: Vinh Nguyen Van (vinhnv@vnu.edu.vn)

Short-Term Memory (Bi-LSTM) networks combined with CRF layers, marked a significant advancement in performance [7]. More recently, pre-trained language models tailored for Vietnamese, such as PhoBERT [6], have achieved state-of-the-art results across various NLP tasks, including NER. The development of domain-specific datasets, notably the VLSP 2016 corpus and the PhoNER COVID-19 dataset [8], has further propelled research in Vietnamese NER, enabling more robust and context-aware entity recognition systems.

Beyond generic NLP applications, Named Entity Recognition plays a vital role in many domain-specific systems where accurate entity extraction is essential such as biomedical text mining, legal document processing, and more recently, e-commerce logistics. However, the study face low-resource languages in specific domains including: (1) the small training data; (2) a large number of labels (hundreds of labels) in the NER system; and (3) domain contexts that differ from the general domain.

Although large language models (LLMs) have become increasingly popular for a wide range of NLP tasks, applying them in specialized, high-stakes domains still poses major challenges. As several studies have noted, LLMs often struggle to capture domain-specific terminology and may produce inconsistent results for rare or technical entities, which limits their reliability in real-world applications such as e-commerce [9]. In this context, NER contributes not only to the automated understanding of user-generated content but also to improving data interoperability and supporting downstream tasks like recommendation, fulfillment tracking, and customer support automation [4].

However, for Vietnamese e-commerce and logistics or specific domain, existing NER models have not been carefully researched and developed because it meets low-resource languages. Most publicly available systems focus on generic entities and fail to capture the fine-grained, context-specific entities critical for logistics workflows such as identifying relationships among buyers, couriers, and dispatch coordinators. This gap creates significant barriers for enterprises aiming to automate and scale operations that rely heavily on informal, short-text messages exchanged across stakeholders.

Within the scope of this research, low-resource Named Entity Recognition (NER) in domain-specific Vietnamese settings is defined as a scenario characterized by three concurrent constraints: limited annotated training data, a large and fine-grained label set, and a substantial distribution shift between general-domain and specialized-domain text. Under such conditions, purely neural approaches often struggle with label sparsity and unreliable generalization, while purely rule-based systems lack flexibility in capturing contextual and linguistic variation.

To address these challenges, a hybrid neurosymbolic architecture is proposed, which integrates rule-based reasoning and deep learning within a unified processing pipeline. In this architecture, a rule-based component is first applied to encode structured domain knowledge and reduce label complexity by grouping related entity categories. A neural sequence labeling model is then trained on the reduced label space to capture contextual patterns more effectively in data-scarce settings. Finally, a post-processing module restores the original fine-grained labels, ensuring both expressive power and practical usability in real-world applications.

The contributions of this research are presented through several key advancements in the field of Natural Language Processing for low-resource languages. Initially, an efficient hybrid Named Entity Recognition (NER) method is proposed for specialized domains, integrating deep learning architectures with rule-based approaches to effectively capture the linguistic complexities of Vietnamese. This framework specifically addresses the challenges inherent in tasks involving a high volume of labels coupled with limited training data. By strategically selecting groups of labels with related properties to reduce initial complexity, the system utilizes designed rules to determine precise labels, thereby streamlining the recognition process in data-scarce environments.

Furthermore, a scalable data augmentation strategy is introduced, enabling the expansion of label sets without the exhaustive requirement for re-annotating existing datasets. This methodology allows for flexible adaptation to evolving operational requirements while simultaneously maintaining model performance and annotation efficiency. To ensure practical applicability, the study incorporates inference-time optimization techniques and post-processing modules. These enhancements significantly improve processing speed and ensure the feasibility of deploying such models within enterprise-grade systems where real-time performance is critical.

The validity of the proposed approach is established through extensive empirical validation on diverse datasets, including real-world logistics, the rare Wildlife dataset, and PhoNER_Covid19. The experimental results demonstrate strong performance across these specialized domains, proving that the hybrid framework remains robust even when faced with the dual constraints of high label density and minimal training samples. This comprehensive suite of developments offers a significant contribution to the development of reliable and scalable NER systems for specialized Vietnamese contexts.

## 2. RELATED WORK

### 2.1. Overview of Named Entity Recognition (NER)

Named Entity Recognition (NER) stands as a fundamental and extensively studied task within the broader field of Natural Language Processing (NLP) [1]. It is formally defined as the process of identifying and classifying predefined categories of named entities in unstructured text, such as proper names of persons, organizations, locations, as well as numerical and temporal expressions [2], [10]. The primary objective of NER is to assign a semantic label to sequences of tokens that refer to specific real-world objects or concepts, thereby transforming raw textual data into a more structured and machine-interpretable format [11]. This structured information is crucial for a multitude of downstream NLP applications, including, but not limited to, information retrieval, where NER helps to refine search queries and improve the relevance of results [12]; question answering systems, by identifying key entities in both questions and potential answer passages [13]; machine translation, by ensuring the correct handling of proper nouns [14]; and knowledge base population, where extracted entities and their relations form the building blocks of structured knowledge graphs [15]. Despite significant advancements, particularly with the advent of deep learning models, NER remains a complex challenge due to linguistic ambiguities, the diversity of entity types, and the inherent variability across different languages and domains [4], [16].

### 2.2. Low-Resource Named Entity Recognition

Low-Resource Named Entity Recognition often occurs when apply it to a specific domain, which has small training data and a large number of labels. The general-purpose NER systems, while achieving considerable success on news articles and general web text, often exhibit a significant performance degradation when applied to domain-specific corpora [1], [17]. This is primarily due to the unique characteristics of specialized domains, such as the biomedical field, legal documents, financial reports, or scientific literature. In the biomedical domain, NER systems are tasked with identifying entities like genes, proteins, diseases, and chemicals. The complexity and specialized vocabulary inherent in biomedical texts pose significant challenges. Crichton et al [17] introduced a multi-task learning approach using convolutional neural networks to improve NER performance across various biomedical datasets. Similarly, Wang et al [18] proposed a deep multi-task learning framework that leverages shared representations to enhance cross-type biomedical NER.

The legal domain presents its own set of challenges for NER, including the identification of legal terminologies, case references, and statutes. Pais et al [19] developed a NER system tailored for the Romanian legal domain, utilizing the LegalNERo corpus and combining multiple word embeddings trained on legal texts. Furthermore, Naik et al [20] conducted an experimental study focusing on NER approaches for Indian legal documents, highlighting the importance of domain-specific datasets and models. In the financial sector, NER is employed to extract entities such as company names, financial instruments, and monetary values from unstructured data sources. Shah et al [21] introduced FiNER-ORD, a high-quality English Financial NER Open Research Dataset, serving as a benchmark for evaluating NER models in the financial domain.

Recently, prompting with LLMs method offers strong generalization capabilities, it proves impractical in the context due to the small of the dataset with a large labels. Prompting methods are computationally expensive, slow in inference time, and ill-suited for fine-grained token-level predictions required in NER. These limitations make them unsuitable for high-throughput, cost-sensitive environments like ours. Additionally, LLM-based prompting approaches also suffer from limited domain adaptability. As highlighted in [9], LLMs often fail to accurately capture domain-specific terminology and contextual nuances, especially in specialized fields such as finance, logistics, or biodiversity. Without extensive domain adaptation or fine-tuning, their predictions tend to be inconsistent, leading to reduced precision and recall when handling rare or technical entities that are critical in the dataset. This further constrains their applicability in real-world, domain-specific NER tasks. These studies across diverse domains highlight a consistent theme: achieving high performance requires domain-specific datasets and tailored models. However, the Vietnamese logistics and e-commerce domain remains significantly under explored, a gap this study aims to address.

## 2.3. Named Entity Recognition for Vietnamese

Named Entity Recognition for Vietnamese (Vietnamese NER) presents a unique set of challenges primarily due to the linguistic characteristics of the language. A major hurdle is the lack of explicit word boundaries, as Vietnamese is an isolating language where words are often composed of single syllables that can also be standalone words. Consequently, word segmentation is a critical and non-trivial preprocessing step, and errors in this stage can significantly propagate and degrade the performance of downstream NER models [22]. Furthermore, Vietnamese exhibits a high degree of ambiguity, where a single word can have multiple meanings or function as both a common noun and part of a proper name, adding another layer of complexity for NER systems [7].

Recently, Named Entity Recognition in Vietnamese has evolved significantly, transitioning from traditional statistical methods reliant on hand-crafted features to sophisticated deep learning architectures, particularly transformer-based pre-trained language models. This progression has been driven by dedicated community efforts, notably through the VLSP shared tasks, which have provided increasingly complex and comprehensive datasets, including those with nested and domain-specific entities. While these resources have advanced general-domain Vietnamese NER, they do not adequately cover the specialized terminologies found in industrial contexts like logistics, where messages are often informal, abbreviated, and context-dependent.

## 2.4. Speed up model

As large language models (LLMs) become increasingly integral to real-time applications, enhancing the inference speed of Transformer-based models has emerged as a critical priority. Recent advancements have introduced a suite of techniques designed to optimize inference efficiency without substantially sacrificing model accuracy, thereby enabling the deployment of LLMs in latency-sensitive contexts.

One cornerstone optimization is Key-Value (KV) Caching, which stores previously computed key and value tensors during autoregressive decoding to eliminate redundant computations. Despite its effectiveness, this method often entails considerable memory overhead. To tackle this limitation, Wu and Tu developed the Layer-Condensed KV Cache, a method that caches KVs for only a select subset of layers[23]. This approach delivers up to 26 times higher throughput compared to standard Transformer models. In a complementary effort, Dai et al proposed CORM, a cache eviction policy that dynamically retains only the most essential key-value pairs[24]. This innovation reduces memory usage by as much as 70%, with minimal impact on performance.

Another breakthrough comes in the form of FlashAttention, which enhances the efficiency of the attention mechanism. By optimizing memory access patterns and minimizing memory reads and writes between GPU high-bandwidth memory and on-chip SRAM, FlashAttention enables faster and more memory-efficient exact attention computations. Tri Dao et al have shown that this technique yields significant speed improvements across both training and inference tasks[25]. Additionally, ONNX Runtime offers a cross-platform framework with built-in optimizations such as operator fusion and quantization, enabling faster Transformer inference. Combined with KV caching and FlashAttention, it contributes to a powerful toolkit for deploying large language models in latency-sensitive applications.

# 3. Methodology

As mentioned above, LLMs are not effective for domain specific NER problems (expensive cost and inference time). Therefore, this study focus on deep learning-based approaches, particularly pre-trained transformer-based models, offering stronger generalization and adaptability across entity categories. They excel at capturing contextual information and handling ambiguous cases, yielding high recall and robustness in diverse scenarios. However, their training and inference cost can be substantial without domain-specific adjustments, and they may overfit frequent patterns while missing rare but critical entities.

To address this problem, this research introduce a hybrid solution that combines rule-based heuristics with fine-tuning of pre-training models. Rule-based components facilitates the exploitation of domain-specific patterns (e.g., phone numbers or timestamps), while fine-tuned transformer-based models provide robust and scalable sequence labeling. The unique point lies in the two-stream parallel processing pipeline, where rule-based entities are filtered out first to reduce the burden on the deep learning model, allowing it to focus on more semantically ambiguous entities. The details of the framework are presented in the following sections.

## 3.1. Problem Define

### 3.1.1. Named Entity Recognition (NER).

Given an input token sequence

$$X = (x_1, x_2, \dots, x_n) \quad (1)$$

the goal of Named Entity Recognition (NER) is to assign an entity label to each token. Formally, NER learns a mapping.

$$f : X \rightarrow Y \quad (2)$$

$Y = (y_1, y_2, \dots, y_n)$ is a predefined set of entity categories. The predicted label sequence is obtained by maximizing the conditional probability.

$$Y = \arg\max_{Y} P( Y \mid X; \theta ) \quad (3)$$

### 3.1.2. Low-Resource Named Entity Recognition

In the standard Named Entity Recognition (NER) setting, a model is trained on a labeled dataset consisting of sequences of tokens, where each token is assigned a corresponding label from a predefined set. The main objective is to learn a mapping that can accurately predict the correct label for each token in a given sequence. During training, the model learns patterns from the data to estimate the likelihood of possible label sequences and selects the most appropriate one for a given input.

Low-resource NER refers to scenarios where the number of annotated samples is extremely limited, or even completely unavailable in zero-shot settings. In such conditions, models cannot rely solely on supervised learning and must instead leverage alternative sources such as unlabeled data or external knowledge. These additional resources may include pretrained language models, domain-specific dictionaries, or cross-lingual information.

Building on this definition, this study focuses on domain-specific Vietnamese NER, where both label scarcity and domain shift pose significant challenges to model performance. In logistics-related NER tasks, the low-resource nature is reflected in the imbalance between the number of samples and the number of entity labels. For example, the CS dataset contains more than 5,800 samples but spans 19 entity types (including the “Other” category), resulting in relatively few examples per class. This imbalance becomes more severe in other datasets: the GAM dataset includes 32 entity types with only around 3,000 samples, while the AI Fluent dataset contains 26 entity types with fewer than 1,200 samples.

Compared to standard NER benchmarks such as Med-NER, these datasets exhibit significantly lower ratios of samples to labels, confirming that they represent genuinely low-resource and highly imbalanced scenarios. In such settings, conventional supervised learning approaches are often insufficient, making it necessary to adopt techniques such as transfer learning or prompt-based adaptation to improve performance.

## 3.2. Proposed Hybrid Framework

Proposed NER framework consists of two component pipelines: model training and inference. These pipelines combine rule-based and deep learning methods to achieve high accuracy and efficiency. It is important to design rules to recognize special labels like dates, IDs, ... and group the labels (reduce the number of labels) then re-detect the original labels. The details of the data annotation pipeline are presented in Section Appendix 1, ensuring a consistent link between the methodology and the datasets used for experimentation.

The NER framework consists of two primary component pipelines: model training and inference. These pipelines combine rule-based and deep learning methods to achieve high accuracy and efficiency. The model training pipeline is designed to handle label sparsity and complexity by integrating rule-based heuristics to pre-process and group entity labels (such as temporal expressions and IDs) before fine-tuning a neural model. This step reduces the label set, making the neural model's task more manageable and improving its ability to generalize across a smaller, imbalanced dataset.

In the inference pipeline, the rule-based component first filters out easily recognized entities (e.g., dates, locations, IDs), reducing the complexity of the label space. The deep learning model then focuses on more complex, semantically ambiguous entities, benefiting from the reduction in label complexity. Afterward, a post-processing step restores the original fine-grained labels, ensuring that the system can recognize all entity types required for the final application.

To ensure a seamless connection between the methodology and the datasets used for experimentation, the data annotation pipeline is described in Appendix 1, where the labeling process is outlined to maintain consistency and quality across the system. This structured approach enables a balance between efficiency in handling large volumes of data and accuracy in recognizing a wide range of entity types, particularly in low-resource settings.

## 3.3. Rule-based Component Design

In this section rules are designed and used for detecting the correct label from a group of labels and patterns that have certain reliability such as time, ID code, ...

Case merge labels: To reduce the number of labels in a domain-specific NER task—an important consideration when training data are limited but label cardinality is high—labels that share common lexical or semantic characteristics and can potentially be decomposed into separate entities after merging are first identified. A two-stage handling strategy is then adopted. During model training and inference, these related labels are aggregated into a single generalized label to simplify the recognition process and improve model stability. Subsequently, in the post-processing phase, a rule-based decomposition step is applied to split the aggregated label into its constituent entities according to predefined linguistic or domain-specific rules. This approach balances model simplicity with fine-grained labeling accuracy.

Identifying Time-Related Labels: Time-related labels are using the SUTime framework, a robust tool proven effective in the industry. SUTime has demonstrated very high accuracy and reliability in recognizing common time-related entities in recognizing time entities in Named Entity Recognition tasks, including dates, times, and durations. This method is directly integrated into the data processing pipeline to ensure high efficiency and precision.

The identification of Citizen Identification Numbers (CINs) is governed by a set of rigorous regulatory criteria. Specifically, the first three digits are required to correspond to a government-defined province code set, ensuring geographical validity. The fourth digit is restricted to the values 0, 1, 2, or 3, as digits outside this range are deemed invalid under the specific regulations established for the 22nd and 23rd centuries. Furthermore, the fifth and sixth digits denote the year of birth, with the additional constraint that if the fourth digit is 2 or 3, the year of birth must be less than 20. In addition to these structural requirements, a contextual validation is performed to ensure that the age associated with the CIN remains within a reasonable demographic range.

In parallel, the identification of order codes is determined by a structured three-part composition. The initial segment functions as a Shop Code to identify the specific retail outlet, followed by a second segment containing routing information for delivery purposes. The final component consists of a random number segment, which must adhere to specific length-dependent constraints. If this segment comprises nine digits, the leading digit cannot be zero; conversely, if the segment extends to ten digits, it must commence with either 0 or 1. This hierarchical validation framework ensures high precision in identifying both personal and logistical entities within the system.

To further clarify the process of merging label groups and mapping them back to the original fine-grained labels, is exemplified within an example from the domain of clothing measurements. In this context, we have labels such as chest circumference (v1), waist circumference (v2), and hip circumference (v3). Initially, these labels are treated independently, but in the rule-based component, they are merged into a unified group, e.g., body measurements, to reduce the label complexity for the deep learning model.

During inference, once the merged group (e.g., body measurements) is identified, the model relies on semantic context and word relationships to distinguish between the original fine-grained labels. For instance, after recognizing the group, the model uses the surrounding words in the text, such as “chest”, “waist”, and “hip”, to correctly map the merged label back to its specific type (v1, v2, or v3). This mapping process ensures that the fine-grained labels are restored accurately, while still benefiting from the simplification provided by the rule-based merging step.

## 3.4. Model Training Pipeline

In the training pipeline, rule-based patterns automatically extract entities that can be deterministically identified, such as phone numbers, IDs, addresses, and temporal expressions. The remaining tokens are tokenized using VN CoreNLP to preserve Vietnamese-specific token boundaries. These filtered tokens are then used to train a deep learning NER model, fine-tuned with speed-up techniques such as quantization, pruning, and ONNX acceleration. The performance of model is validated against a held-out test set. The entire training process is shown in figure 1.

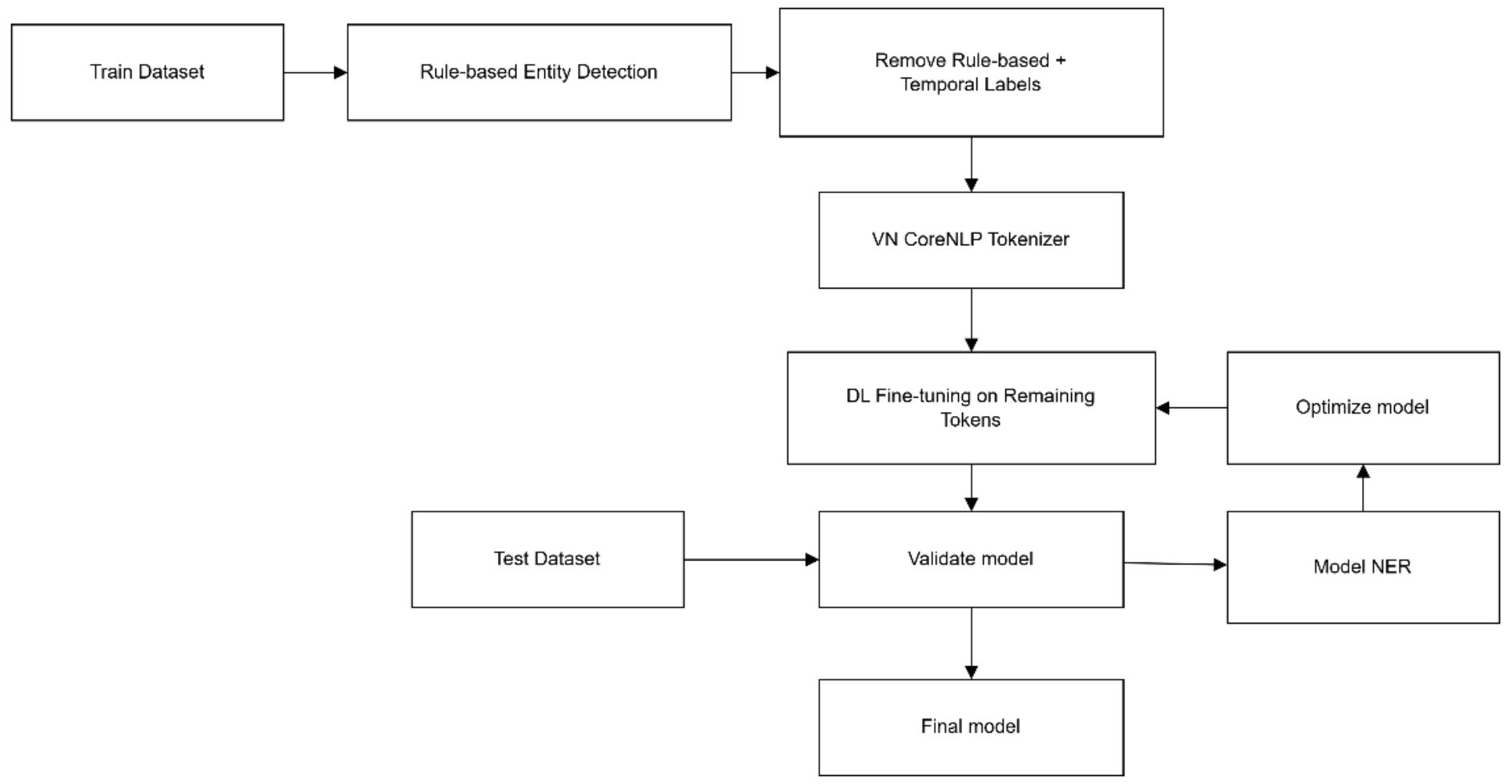

**Figure 1.** Training Phase

## 3.5. Data augmentation

In the data annotation pipeline, raw text is initially labeled using large language models (LLMs) such as ChatGPT, Gemini, or DeepSeek. This paper utilized carefully designed prompts, including Chain-of-Thought, few-shot, and self-consistency techniques, to generate these initial labels. Human annotators then review and correct errors, ensuring a high-quality labeled dataset. This process is illustrated in figure 2.

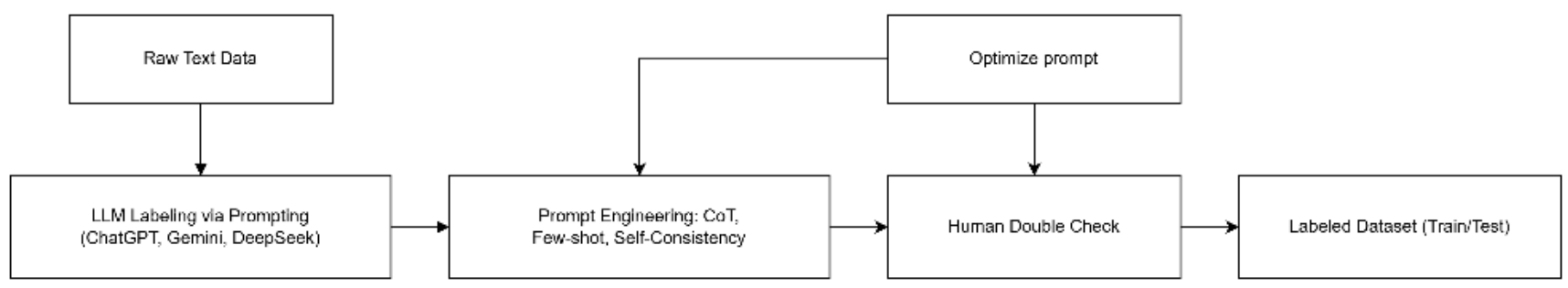

**Figure 2.** Data annotation for all dataset

Data annotation is the foundation of the proposed NER pipeline, combining the rapid processing power of large language models (LLMs) with thorough human verification to ensure both coverage and accuracy. The entire process is illustrated in figure 1.

### 3.5.1. LLM Labeling via Prompting

To enhance performance, particularly within zero-shot and few-shot scenarios, a comprehensive suite of prompt engineering techniques has been implemented. The approach begins with Chain-of-Thought (CoT) prompts, which guide the model through step-by-step reasoning to address complex entities, complemented by few-shot examples that illustrate ideal labeling patterns for context-based learning. To further ensure reliability, Self-Consistency decoding is applied to aggregate multiple outputs, thereby allowing for the selection of the most dependable label. This is reinforced by a Self-Refine mechanism, where the model produces an initial annotation, critically reviews its own output for errors or omissions, and subsequently generates an improved version. This iterative refinement process significantly enhances label accuracy and minimizes boundary inconsistencies.

In addition to individual refinement strategies, a Multi-Agent Debate framework is utilized to foster consensus-based accuracy. In this setup, multiple Large Language Model (LLM) agents independently annotate the same text and compare their outputs through a debate-style exchange. Divergent opinions are reconciled by consolidating overlapping evidence, leading to high-quality consensus labels. This collaborative strategy, drawing upon methodologies such as those by Liu et al. [28], strengthens reasoning and robustness in domain-specific annotation. Collectively, this innovative combination of techniques ensures that the models operate at peak efficiency without relying on extensive manual supervision.

#### 3.5.2. Prompt Optimization

We do not stop there—based on initial results, we continuously adjust prompt templates and example selections. This iterative process, guided by precision and recall metrics on a separate validation set, refines the approach until performance stabilizes, ensuring each modification adds value.

#### 3.5.3. Human Double Check

Every machine-generated label is meticulously reviewed by a team of skilled linguists. They identify errors, resolve ambiguities, and adjust entity boundaries, bringing their expertise to ensure the final dataset meets the highest quality standards.

#### 3.5.4. Labeled Dataset Production

Once validated, the corpus is split into training and test sets. This labeled dataset combines the broad coverage of LLMs with the reliability of human verification, laying a solid foundation for subsequent NER model development.

Together, these steps form an efficient yet thorough labeling framework, balancing speed and accuracy to handle entity labeling across diverse industrial domains.

#### 3.5.5. Value of Data Annotation

To further ensure the reliability of our labeled dataset, we integrate external data sources such as domain-specific knowledge bases and publicly available datasets. This additional layer of validation enhances the trustworthiness of the annotations by cross-referencing with industry standards and expert definitions. By incorporating external data, we reduce the risk of label inconsistencies and biases, ensuring that the labeled dataset is comprehensive, balanced, and aligned with real-world domain knowledge. This approach significantly improves the model's generalizability, accuracy, and adaptability to various industrial domains, such as logistics and healthcare.

#### 3.5.6. Iterative Annotation Framework and Quality Assurance

To ensure the high fidelity of the corpus, a multi-stage, hybrid annotation framework is employed, integrating automated Large Language Model (LLM) generation with human expertise and external validation. This process is iterative rather than static; prompt templates and example selections are continuously refined based on initial outputs. Such iterative refinement is strictly guided by precision and recall metrics evaluated on a dedicated validation set, ensuring that each modification contributes to the stabilization and enhancement of the model's performance.

The transition from machine-generated labels to a finalized gold standard involves rigorous Human-in-the-loop (HITL) verification and external cross-referencing. Every LLM-generated label undergoes meticulous review by skilled linguists to resolve ambiguities, adjust entity boundaries, and correct systematic errors. To further fortify the reliability of these annotations, external data sources—such as domain-specific knowledge bases and publicly available datasets—are integrated as objective anchors. This multi-layered validation reduces the risk of label inconsistencies and biases by aligning the dataset with established industry standards and expert definitions.

In the final production phase, the validated corpus is partitioned into training and test sets. This unified approach, which combines the computational efficiency of LLMs with the nuanced reliability of human expertise and external data, establishes a robust foundation for subsequent Named Entity Recognition (NER) development. The resulting dataset not only exhibits high accuracy but also demonstrates superior generalizability across diverse industrial domains, such as logistics and healthcare, where precision is paramount.

## 3.6. Training method

The models are trained using the cross-entropy loss function, which is computed over all valid token positions. For a sequence of tokens $x = (x_1, x_2, \dots, x_T)$ , the objective of NER is to assign a $y_t$ to each token $x_t$. The output probabilities for labels are computed using a softmax layer:

$$P(y_t = k|x_t) = \frac{exp\big(f_k(x_t,\theta)\big)}{\sum_{k'=1}^{K} exp\big(f_{k'}(x_t,\theta)\big)} \tag{4}$$

where $f_K(x_t;\theta)$ is the activation corresponding to class k, and K is the number of entity classes. The loss function used for training is typically the categorical cross-entropy loss:

$$L = -\sum_{t=1}^{T}\sum_{k=1}^{K} I(y_t = k)P(y_t = k|x_t) \tag{5}$$

(5) where $I(y_t = k)$ is an indicator function that equals 1 if the correct label is k and 0 otherwise.

To build robust neural models for Named Entity Recognition (NER), a sequence labeling framework is adopted in which each input token is assigned a corresponding entity tag. The experiments cover a range of transformer-based language models, including both multilingual and Vietnamese-specific variants. All models are fine-tuned following the strategy illustrated in figure 3.

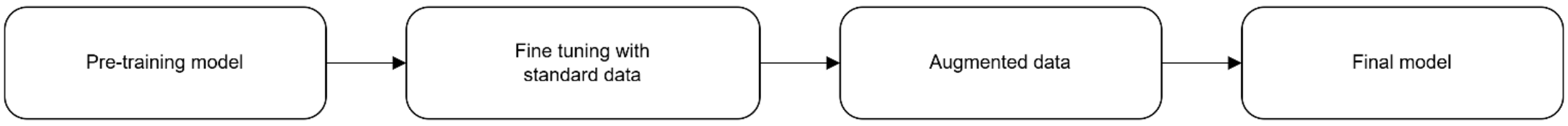

**Figure 3.** Strategy-training

To enhance model generalization and mitigate overfitting, this study further augment the training set with additional synthetic and human-labeled samples. This expansion aims to address data sparsity, particularly for low-frequency entity types. This architecture setup enables a comparative evaluation of both general-purpose and Vietnamese-specific transformers in the context of fine-grained NER.

## 3.7. Inference Phase

During the inference pipeline, the system simultaneously applies rule-based extraction, including SUTime for temporal entities, and the optimized deep learning model. The outputs from these approaches are merged to produce a unified prediction. This hybrid design leverages the precision of handcrafted rules and the generalization ability of neural models, delivering robust NER performance in both high-recall and low-latency scenarios. The inference process is depicted in figure 4.

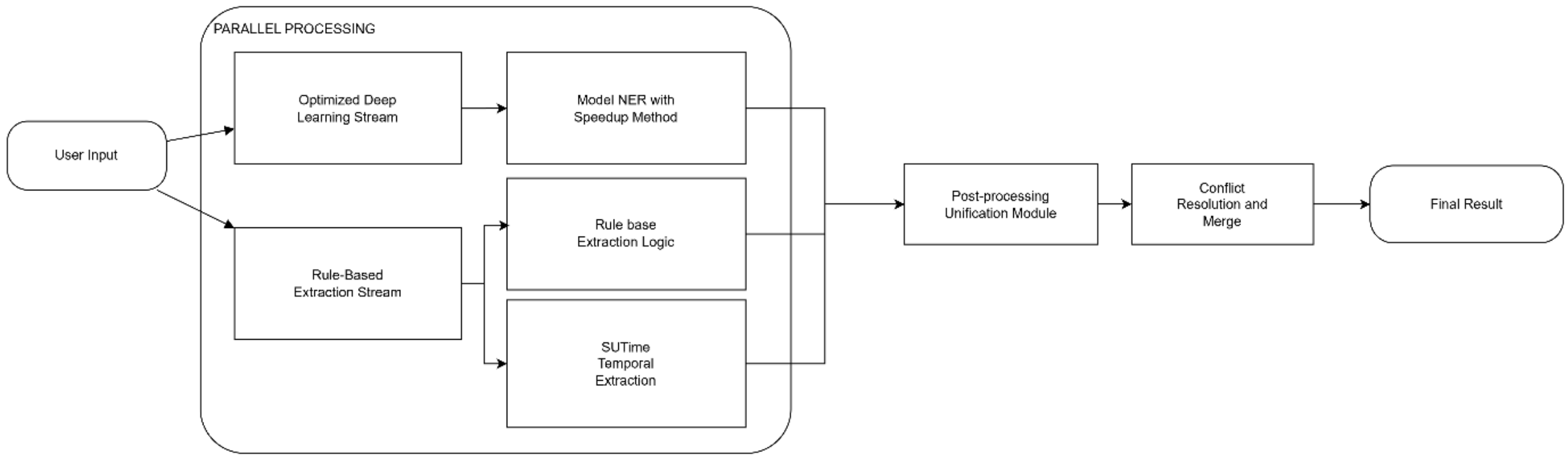

**Figure 4.** Inference for System

Following the generation of raw outputs by the Named Entity Recognition (NER) model, a post-processing stage is implemented to rectify systematic segmentation and labeling errors. A primary challenge addressed during this phase is the occasional assignment of labels to partial spans or sub-tokens—for instance, the model might only identify "Pham" within "Pham Minh Chinh" or "Vietcom" in "Vietcombank." To resolve such discrepancies, a span-reconstruction mechanism is utilized. Whenever a labeled token is identified as part of a larger linguistic or domain-specific expression, lexical and domain-specific rules are applied to re-segment the entire phrase. The identified label is then propagated across the full segment, ensuring total consistency throughout the entity span.

In cases where overlapping spans occur, a merging strategy is employed that prioritizes the longest candidate when multiple tokens share the same label type. Furthermore, to mitigate instances where the model might erroneously merge multiple adjacent entities into a single complex unit, a rule-based splitting procedure is integrated. This procedure systematically identifies and separates these merged entities into their original constituent components, ensuring the final output maintains the necessary granularity.

Collectively, these strategies significantly reduce under-segmentation errors and enhance boundary consistency, resulting in refined annotations that align more closely with domain-specific requirements. By enforcing such span-level coherence, the final entity annotations achieve a higher degree of robustness, which is essential for the accuracy of downstream applications such as information extraction and the construction of complex knowledge graphs.

### 3.8. Speed-up Strategy

In production environments, Named Entity Recognition (NER) systems are required to ingest and label millions of records daily, a volume that significantly exceeds the capacity of standard inference pipelines. Profiling reveals that transformer inference constitutes the primary source of latency; consequently, a three-stage acceleration strategy has been developed to mitigate these bottlenecks and enhance throughput.

Central to this approach are inference-level optimizations, specifically the integration of Key-Value (KV) caching to store previously computed pairs for reuse in subsequent steps. By eliminating redundant attention computations, the overall processing cost is markedly reduced. This efficiency is further augmented by the incorporation of FlashAttention, which optimizes the self-attention mechanism by minimizing memory traffic and overhead, thereby maximizing GPU throughput. To achieve peak execution speeds, models are exported to the Open Neural Network Exchange (ONNX) format and executed via the ONNX Runtime, leveraging graph-level fusions and quantized kernels.

Beyond neural architectural refinements, hardware-level parallelism and dynamic workload management are employed to ensure sustained system efficiency. All non-neural tasks, including rule-based temporal tagging and regex-driven entity matching for identifiers such as phone numbers or product codes, are executed on dedicated CPU threads. This architecture decouples lightweight preprocessing and post-processing from the GPU-intensive inference phase, allowing both components to operate at maximum capacity. Complementing this is a dynamic batching mechanism, where incoming records are buffered and grouped into variable-sized batches according to arrival rates and sequence lengths. This adaptive approach minimizes padding overhead and optimizes GPU utilization in real-time.

Through the synergy of these optimizations, a substantial reduction in end-to-end latency is achieved, decreasing from approximately 50 ms to 13 ms per record. This 75% improvement in performance enables the system to satisfy strict service-level objectives while maintaining the high degree of fine-grained accuracy essential for industrial-scale NER deployments.

### 3.9. Experiments

To verify the effectiveness and comprehensiveness of the proposed method, this research evaluated the framework acrossed five diverse NER datasets in Vietnamese specific domain, each reflecting different industrial or societal contexts where precise entity extraction is critical. Specifically, the first 3 datasets are real customer data provided by a logistics company. A brief description of each dataset is provided below, while detailed statistics and annotation schemes are deferred to Appendix 1.

### 3.9.1. Datasets

#### 3.9.1.1. Customer Service (CS)

This dataset consists of real-world customer interactions collected from support channels such as live chat logs, email threads, and ticketing platforms. It captures 18 fine-grained entity types, including transactional identifiers (e.g., Order Code), personal information (e.g., Phone Number, Customer Name), and logistics details (e.g., Address, Bank Name). The CS dataset underpins applications in automated case routing and service-level analytics, making it representative of high-volume, text-heavy enterprise workflows.

#### 3.9.1.2. General Attribute Modeling (GAM)

The GAM dataset integrates structured profiles with free-text user preference statements from large-scale e-commerce surveys. With 36 annotated entity types, ranging from concrete product metadata (e.g., Product Name, Quantity, Price) to personal attributes (e.g., Skin Type, Body Shape, Fashion Style), this resource reflects the complexity of personalization in modern retail. It provides a foundation for attribute-based recommendation and dynamic product card generation.

#### 3.9.1.3. AI Fluent

Drawn from HR and payroll documentation, this dataset captures semi-structured corporate text including salary slips, allowance forms, time sheets, and policy updates. It spans 26 entity types covering both financial (e.g., Salary, Allowance, Payment Method) and temporal dimensions (e.g., Work Hours, Time Range, Payment Period). The dataset supports downstream payroll reconciliation and compliance monitoring, where accuracy is particularly crucial due to regulatory constraints.

#### 3.9.1.4. PhoNER_COVID19

PhoNER_COVID19 is a domain-specific Vietnamese corpus curated for pandemic-related text analytics [...]. It contains over 10K sentences annotated with 10 entity categories such as PATIENT_ID, SYMPTOM, TRANSPORTATION, and LOCATION. While originally developed for COVID-19 monitoring, the schema generalizes broader epidemic response tasks, offering valuable benchmarks for healthcare NER in low-resource languages.

#### 3.9.1.5. Rare Wildlife

This dataset is derived from Vietnamese news reports on endangered species and conservation efforts. It encodes six entity categories central to biodiversity monitoring: SPECIES, LOCATION, HABITAT, ID_FEATURE, ORGANIZATION, and DATE. Despite its smaller size compared to other corpora, Rare Wildlife provides a unique testbed for ecological NLP, emphasizing multi-word entity recognition in a domain where timely extraction can directly inform conservation policy.

### 3.9.2. Evaluation Metric

To evaluate the performance of frame detection model formulated as a Named Entity Recognition (NER) task This research adopted the standard metrics in sequence labeling: Precision, Recall, and F1 score. These metrics offer a comprehensive view of the model’s ability to correctly identify entity spans and their associated types within unstructured text.

Let TP denote the number of true positives, i.e., entity spans that are correctly predicted with both correct boundaries and correct entity types. Let FP denote false positives predicted spans that do not match any gold annotation and FN denote false negatives gold spans missed by the model. The evaluation metrics are computed as follows:

$$precision = \frac{TP}{TP + FP} \tag{6}$$

$$recall = \frac{TP}{TP + FN} \tag{7}$$

$$F1Score = \frac{2 \times \Pr e\,cision \times Recall}{\Pr e\,cision + Recall} \tag{8}$$

The report entity-level metrics, meaning that a prediction is considered correct only if both the span boundaries and the entity type match the ground truth exactly. This is a more stringent criterion than token-level evaluation and is more appropriate for downstream applications that rely on precise entity extraction.

### 3.9.3. Configure setup

#### 3.9.3.1. Data

To evaluate the effectiveness of various pre-trained Transformer-based models on Vietnamese Named Entity Recognition (NER), This study conducted experiments on five distinct datasets: Customer Service, GAM, AI-Fluent, Pho NER covid19 [8] and Rare Wildlife.

Each dataset was preprocessed follow train and validated dataset. The label sets across the three datasets are not fully aligned, which further highlights the need for models that can adapt to varying entity types and naming conventions. All datasets used in the experiments are described in detail in Section Dataset.

#### 3.9.3.2. Tools

Although production environment includes multiple servers to distribute inference workloads at scale, all experimental results reported in this paper were obtained using a single server equipped with an NVIDIA RTX 3090 Ti GPU (24GB VRAM) for both training and evaluation information server in table 1. This constraint simulates realistic resource-limited research settings and establish a reproducible baseline for latency and throughput measurements.

**Table 1.** System Hardware Configuration Used for Model Training and Infer

| Component | Specification |
|---|---|
| GPU | NVIDIA RTX 3090 Ti (24GB VRAM) |
| CPU | Intel(R) Xeon(R) Silver 4114 @ 2.20GHz |
| RAM | 128GB DDR4 ECC |
| Storage | 2TB NVMe SSD |
| Operating System | Ubuntu 22.04 LTS (64-bit) |
| CUDA Version | 11.8 |
| Driver Version | 525.60.13 |

By standardizing all experiments on a single machine, this study are able to reliably estimate the inference cost per record, which in turn to extrapolate the expected total inference time across the full production database. This setup provides actionable insights for deployment planning, capacity estimation, and cost-performance trade-off analysis in real-world industrial systems.

#### 3.9.3.3. Hyper Parameters

The experiment optimized model parameters by using the Adam optimizer [27]. Additional training hyperparameters such as learning rate, batch size, and epoch count will be detailed in table 2.

**Table 2.** Hyperparameter Settings for Transformer-based NER Models

| Parameter | BERT | PhoBERT | BERT-Large |
|---|---|---|---|
| Learning Rate | 2e-5 | 2e-5 | 3e-5 |
| Batch Size | 32 | 32 | 16 |
| Max Length | 256 | 256 | 512 |
| Warmup Steps | 500 | 500 | 1000 |
| Weight Decay | 0.01 | 0.01 | 0.01 |
| Number of Epochs | 10 | 10 | 10 |
| Optimizer | Adam | Adam | Adam |
| Scheduler | Linear Decay | Linear Decay | Linear Decay |
| Dropout Rate | 0.1 | 0.1 | 0.1 |

#### 3.9.3.4. Training Scenario

The training pipeline follows a three-stage procedure illustrated in figure 3. A pre-trained language model is first adopted and subsequently subjected to two sequential fine-tuning phases: (1) fine-tuning with standard labeled data and (2) fine-tuning with augmented data to enhance generalization.

Phase 1: Fine-tuning with Standard Data. Each model is initialized from publicly available pre-trained checkpoints. In the first phase, the models are fine-tuned using the original NER datasets described in Appendix. This phase enables the models to adapt their representations to the Vietnamese NER task without incorporating domain-specific augmentation.

Phase 2: Fine-tuning with Augmented Data. Following the initial fine-tuning with standard annotated data, a second training phase is conducted using augmented data. The objective is to improve model generalization and robustness, particularly in handling rare or noisy entity mentions. The augmentation techniques include synonym replacement, entity span recombination, and back-translation, all of which are designed to preserve entity integrity while diversifying contextual patterns.

Two training strategies are explored for this phase: (1) training solely on the augmented data, and (2) merging the standard and augmented datasets into a single training set. Empirical results indicate that separating the phases—first fine-tuning on the original data and then continuing training on the augmented data—yields better generalization and more stable entity boundary predictions. In contrast, merging both datasets from the beginning tends to introduce noise early in training, which can hinder convergence and reduce final F1 scores.

Therefore, in final setup, To adopt the sequential approach where augmented data is used in a distinct second phase, following standard data fine-tuning. This design choice leads to more consistent improvements across all model variants evaluated.

To ensure a rigorous evaluation of Named Entity Recognition (NER) performance, particularly within the context of the Vietnamese language, six Transformer-based models with distinct architectures and pretraining strategies are analyzed. These models represent the state-of-the-art in both general linguistic representation and language-specific optimization. The selection begins with BERT [3], the foundational bidirectional transformer pretrained on English corpora, and RoBERTa [26], an optimized variant that improves upon the original BERT through dynamic masking and extended training durations. To address the specificities of the Vietnamese language, PhoBERT [6] is included, as it was pretrained on large-scale monolingual Vietnamese datasets to capture local semantic nuances.

The evaluation further explores advanced attention mechanisms through the DeBERTa architecture, incorporating both mDeBERTa and ViDeBERTa. The former is a multilingual model designed to capture complex dependencies using disentangled attention and enhanced positional encoding, while the latter provides a version specifically localized through Vietnamese pretraining. Additionally, ViT5, a sequence-to-sequence pretrained transformer based on the T5 architecture and trained on Vietnamese text, is evaluated to assess the efficacy of encoder-decoder frameworks in entity labeling tasks.

To maintain a fair and consistent comparison, all models are trained using a standardized optimization configuration. This includes the implementation of the AdamW optimizer and a linear learning rate scheduler with a warm-up phase. To prevent overfitting, an early stopping mechanism is employed based on the validation F1 score. The experimental workflow is supported by the Hugging Face transformers library for model deployment, while the seqeval framework is utilized for the standardized evaluation of sequence labeling performance. Hyperparameter follow table 2 and Evaluation Metric follow equation Precision, Recall and F1-Score.

## 4. Results And Discussion

Table 3, table 4, and table 5 summarize the results obtained across the three tasks. As shown in table 3, both PhoBERT and RoBERTa achieved the highest performance on the customer service dataset, each reaching an F1 score of 89.7. This indicates their strong ability to capture domain-specific entities, possibly due to their robust pretraining on Vietnamese corpora (PhoBERT) and deeper contextual modeling (RoBERTa).

**Table 3.** System Performance on the Customer Service Task

| Model | Precision | Recall | F1 Score |
|---|---|---|---|
| BERT | 80.24 | 77.55 | 79.38 |
| BERT + Post-proc. | 81.02 | 78.22 | 80.10 |
| PhoBERT | 88.91 | 90.12 | 89.51 |
| PhoBERT + Post-proc. | 89.76 | 90.84 | 90.27 |
| ViT5 | 84.67 | 85.34 | 84.95 |
| ViT5 + Post-proc. | 85.42 | 86.02 | 85.71 |
| BERT Large | 82.88 | 84.11 | 83.49 |
| BERT Large + Post-proc. | 83.55 | 84.92 | 84.22 |
| RoBERTa | 89.34 | 90.07 | 89.70 |
| RoBERTa + Post-proc. | 90.02 | 90.89 | 90.46 |
| mDeBERTa | 78.23 | 76.41 | 77.31 |
| mDeBERTa + Post-proc. | 78.94 | 77.12 | 78.03 |

ViT5 and BERT Large followed closely with F1 scores of 84.95 and 89.51 respectively, suggesting that model size and multilingual capability contribute positively, but not as significantly as domain adaptation. Surprisingly, mDeBERTa, despite being a highly efficient multilingual model, underperformed with an F1 score of only 77.31, possibly due to a domain mismatch or insufficient fine-tuning on the customer service context.

**Table 4.** System Performance on the GAM Task

| Model | Precision | Recall | F1 Score |
|---|---|---|---|
| BERT | 70.32 | 75.84 | 73.05 |
| BERT + Post-proc. | 71.02 | 76.41 | 73.68 |
| RoBERTa | 81.27 | 87.93 | 84.48 |
| RoBERTa + Post-proc. | 82.01 | 88.66 | 85.09 |
| BERT Large | 73.18 | 76.22 | 74.67 |
| BERT Large + Post-proc. | 73.89 | 77.01 | 75.32 |

The AI Fluent task (table 5) saw mDeBERTa rise to the top with an F1 score of 83.54, closely followed by RoBERTa 81.67 and PhoBERT 80.75. This pattern reflects the benefits of multilingual and deep models when applied to complex or semi-structured entity tasks such as AI-generated responses. ViT5 and BERT Large both performed competitively 79.84, while BERT lagged behind 73.46, likely due to limited capacity in modeling long-range dependencies across varied entity types. It is noteworthy that mDeBERTa, which previously underperformed on the customer service task, generalized better in this scenario—implying that its architecture may benefit more from syntactically richer or longer-form input like those in the AI Fluent dataset.

**Table 5.** Example of full column size table

| Model | Precision | Recall | F1 Score |
|---|---|---|---|
| BERT | 72.14 | 74.82 | 73.46 |
| BERT + Post-proc. | 72.88 | 75.51 | 74.18 |
| BERT Large | 80.07 | 79.95 | 80.01 |
| BERT Large + Post-proc. | 80.83 | 80.62 | 80.73 |
| PhoBERT | 78.92 | 82.67 | 80.75 |
| PhoBERT + Post-proc. | 79.64 | 83.41 | 81.49 |
| ViT5 | 77.83 | 81.92 | 79.84 |
| ViT5 + Post-proc. | 78.59 | 82.61 | 80.58 |
| RoBERTa | 80.25 | 83.12 | 81.67 |
| RoBERTa + Post-proc. | 81.03 | 83.89 | 82.46 |
| mDeBERTa | 82.17 | 84.95 | 83.54 |
| mDeBERTa + Post-proc. | 82.94 | 85.63 | 84.28 |

Across all three tasks, RoBERTa demonstrated consistently high performance, affirming its general robustness across domains. PhoBERT excelled in Vietnamese-specific contexts, particularly customer service. mDeBERTa, while initially underperforming, showed significant strength in AI Fluent, highlighting the importance of task–model alignment. Overall, these results emphasize the trade-offs between domain-adapted models (e.g., PhoBERT), general multilingual transformers (e.g., mDeBERTa), and model depth (e.g., BERT Large vs. BERT base), which should guide model selection based on task characteristics.

## 4.1. Application to the Medical and Rare Wildlife NER Task

Data processing strategy is applied to the Named Entity Recognition (NER) task in the medical domain, using the publicly available Vietnamese COVID-19 NER dataset [8]. As shown in figure 3, this strategy leads to improved model performance compared to baseline approaches. As shown in table 7, the strategy consistently improves the F1-score across all models.

**Table 6.** Comparison of model performance on Vietnamese COVID-19 NER dataset with and without data strategy

| Model | Precision | Recall | F1-Score |
|---|---|---|---|
| Bert-base | 81.10 | 79.88 | 80.48 |
| Bert-base (Our strategy) | 84.30 | 83.15 | 83.72 |
| Roberta-base | 84.50 | 83.50 | 84.00 |
| Roberta-base (Our strategy) | 90.80 | 89.60 | 90.18 |
| Roberta-large | 92.20 | 91.42 | 91.81 |
| Roberta-large (Our strategy) | 94.60 | 93.70 | 94.15 |

**Table 7.** Comparison of model performance on Rare Wildlife NER dataset with and without data strategy

| Model | Precision | Recall | F1-Score |
|---|---|---|---|
| Bert-base | 81.10 | 79.88 | 80.48 |
| Bert-base (Our strategy) | 84.30 | 83.15 | 83.72 |
| Roberta-base | 84.50 | 83.50 | 84.00 |
| Roberta-base (Our strategy) | 90.80 | 89.60 | 90.18 |
| Roberta-large | 92.20 | 91.42 | 91.81 |
| Roberta-large (Our strategy) | 94.60 | 93.70 | 94.15 |

Four inference acceleration techniques—Standard, Flash Attention, ONNX Runtime, and KV Cache—on a sequence generation model using 9,000 tokens as input, deployed on NVIDIA Tesla P100 (Kaggle). As shown in table 8, KV Cache achieved the highest speed at 1,792.8 tokens/s, a 5× improvement over standard decoding. Flash Attention and ONNX also provided significant speedups (3.4× and 2.9× respectively), with no increase in GPU memory usage. All methods preserved output quality, confirming that acceleration was achieved without compromising correctness.

**Table 8.** Comparison of inference acceleration techniques

| Method | Tokens | Time(s) | Tokens/s | Hardware | GPU Used |
|---|---|---|---|---|---|
| Standard Inference | 9000 | 25.3 | 355.7 | P100 (Kaggle) | 2.4 GB |
| Flash Attention | 9000 | 7.49 | 1201.6 | P100 (Kaggle) | 2.4 GB |
| ONNX | 9000 | 8.64 | 1041.6 | P100 (Kaggle) | 2.4 GB |
| KV Cache | 9000 | 5.02 | 1792.8 | P100 (Kaggle) | 2.4 GB |

## 4.2. Analysis

The observed variations in performance across the NER tasks stem from inherent differences in task complexity and the linguistic properties of each domain. The Customer Service (CS) task, involving entities such as product names and addresses, aligns closely with everyday Vietnamese language patterns. This alignment renders the task less ambiguous and more predictable, enabling models to achieve high F1-scores of up to 90%, with rule-based methods proving particularly effective due to the well-defined nature of its entity set.

In contrast, the General Attribute Modeling (GAM) and AI Fluent tasks pose greater challenges due to their domain-specific demands. These tasks require a nuanced grasp of contextual and semantic relationships, as their fine-grained labels transcend simple lexical recognition. For instance, the GAM task necessitates distinguishing subtle attribute variations, while the AI Fluent task incorporates dense business and workflow terminology. This added complexity results in slightly lower peak F1-scores 85% for GAM with a RoBERTa-based model and 84% for AI Fluent reflecting the increased difficulty of modeling such specialized domains.

A similar trend is observed in the PhoNER Covid-19 dataset (table 6), where domain-specific terminology and emerging entity types such as case numbers, symptoms, and administrative regions introduce moderate complexity. Nevertheless, thanks to the relatively larger availability of annotated data, models achieve high performance, with RoBERTa-large reaching an F1-score of 91.81%, further boosted to 94.15% with data strategy. The consistent gains across all backbones highlight that augmentation approach not only strengthens generalization but also mitigates overfitting in specialized but data-rich domains.

By contrast, the Rare Wildlife NER dataset presents the most challenging scenario (table 6). Here, entities correspond to rare and low-frequency species names, leading to extreme data sparsity and lexical variability. Baseline models such as RoBERTa-base and RoBERTa-large achieve only around 36 F1, underscoring the difficulty of modeling this under-resourced domain. With data strategy, however, both models exhibit dramatic improvements of over 24 points in F1-score, primarily due to the substantial increase in training data coverage. This demonstrates that data augmentation is particularly crucial in low-resource settings, where the lack of sufficient training examples severely hampers recognition performance. In addition to dataset size and label complexity, another key factor in the performance improvement is the high class imbalance within the dataset. Many entity categories are rare, causing traditional NER models to overfit on the more frequent labels while struggling with less frequent ones. The hybrid approach is particularly well-suited to handle this imbalance, as the rule-based components reduce label complexity by grouping similar entity types, while the fine-tuned pre-trained models generalize better across the smaller, imbalanced dataset.

The low baseline performance (F1 scores ranging from 36% to 60%) can be attributed to both the small sample size and label sparsity. Many of the entity types are rare, which makes it challenging for standard models to learn robust representations. Furthermore, these entities often consist of multi-word spans, adding complexity to accurate boundary matching. These factors lead to poor generalization and overfitting on more frequent labels. The hybrid approach, combining rule-based heuristics and fine-tuned pre-trained models, effectively addresses these challenges by reducing label complexity and enabling the model to learn more effectively from limited data.

The consistent outperformance of hybrid approaches, which integrate rule-based techniques with multilingual deep learning models, highlights the value of combining explicit linguistic knowledge with data-driven learning. Rule-based methods excel at capturing entities with predictable patterns, whereas deep learning models adeptly handle context-dependent entities. This synergy enhances system robustness, particularly in low-data scenarios or when entities are clearly delineated, explaining the observed performance advantage over standalone deep learning systems.

Turning to inference performance, the substantial speed improvements from optimization techniques Flash Attention, ONNX conversion, and KV Caching arise from their ability to address distinct computational bottlenecks. Flash Attention reduces inference time to 7.49 seconds (a 3.38x speedup) by restructuring the attention mechanism to minimize memory reads/writes and optimize GPU computation, proving especially effective for the long sequences tested (9000 tokens). ONNX conversion, achieving 8.64 seconds (a 2.93x improvement), optimizes the model graph for hardware-specific acceleration, enhancing overall runtime efficiency. Most strikingly, KV Caching cuts inference time to 5.02 seconds (a 5.04x speedup) by storing intermediate key and value matrices, eliminating redundant computations during autoregressive decoding.

These findings carry significant implications for designing efficient and effective NLP systems. The analysis underscores the necessity of tailoring solutions to both the linguistic characteristics of the task and the computational demands of the model. By elucidating the reasons behind performance variations, this study informs strategic choices in model architecture, optimization strategies, and the incorporation of domain knowledge, paving the way for advancements in real-time and large-scale NLP applications.

A direct empirical comparison with LLM-based prompting was not conducted due to significant computational costs. However, the specialized fine-tuned model is inherently more efficient for high-throughput industrial applications, where the documented high latency and cost of LLMs would be prohibitive. The system struggles to accurately recognize nested or compound named entities, particularly when an entity consists of both a functional descriptor (e.g., "Ngân hàng Ngoại thương") and a brand name or abbreviation (e.g., "Vietcombank"). This results in partial recognition or complete omission of the Ngân hàng (Bank) label.

When multiple bank names are listed together in a coordinated structure (e.g., "Agribank, VietinBank, BIDV"), the system often fails to segment and label them individually. This indicates a limitation in handling entity enumeration patterns, which are common in customer service and financial documents. Despite the clear presence of context indicating a financial institution (e.g., "dịch vụ đặt mua vàng", "website của ngân hàng"), the model sometimes fails to leverage this context to reinforce or disambiguate entity classification, revealing a weakness in contextual consistency and coreference grounding (table 9).

**Table 9.** Examples in Named Entity Recognition for the Customer Service Task

| Sentence | Expected Entity | Predicted Output | Error Type |
|---|---|---|---|
| Ngân hàng Ngoại thương (Vietcombank) thông báo dừng cung cấp dịch vụ đặt mua vàng miếng trên website từ hôm nay. (The Joint Stock Commercial Bank for Foreign Trade of Vietnam (Vietcombank) announced it would stop providing gold bar purchase services on its website from today.) | <bank>Vietcombank</bank> | Not predicted | Missed nested entity |
| 3 nhà băng quốc doanh khác là Agribank, VietinBank, BIDV và Công ty Vàng bạc đá quý Sài Gòn (SJC)... (3 other state-owned banks are Agribank, VietinBank, BIDV and Saigon Jewelry Company (SJC)...) | <bank>Agribank, VietinBank, BIDV</bank> | <bank>Agribank</bank> | List entity recognition failure |
| Khách hàng có thể đặt mua vàng qua BIDV hoặc Vietcombank từ hôm nay. (Customers can order gold through BIDV or Vietcombank starting today.) | <bank>BIDV, Vietcombank</bank> | <bank>BIDV, Vietcombank</bank> | No error |
| Dịch vụ này chỉ áp dụng trên ứng dụng ngân hàng điện tử. (This service only applies to the digital banking application.) | No label | <organization>digital banking application</organization> | Overlabeling / hallucination |
| Công ty Vàng bạc đá quý Sài Gòn (SJC) sẽ hỗ trợ xác nhận đơn hàng trực tuyến. (Saigon Jewelry Company (SJC) will support online order confirmation.) | <organization>SJC </organization> | <organization>SJC </organization> | No error |

GAM dataset has some errors (see table 10): Incomplete Span Detection: The system often fails to capture the full range of entities, especially when address entities are involved. For example, it often truncates location names by omitting suffixes such as districts or cities (e.g., "tòa nhà Avalon, quận 1 is labeled only as "tòa nhà Avalon"), severely affecting entity utility for downstream tasks like geolocation or case resolution.

Missed Entities with Clear Indicators: Even when tokens or phrases strongly indicate a location (e.g. “Cảng Sao Mai”, “tỉnh Bà Rịa - Vũng Tàu”)}, the model sometimes fails to assign any label. This indicates weaknesses in recognizing structured but informal place references or specialized named locations. Partial numerical extraction: The model inconsistently extracts product or customer measurements (e.g., only identifying "2,16 m" but missing "1,2 m"), which hampers its use in e-Commerce and logistics domains that require precise dimension tracking. Inconsistent Multi-word Unit Recognition: Phrases such as "vàng miếng are occasionally only partially labeled (e.g., labeling just "miếng), suggesting an issue with recognizing multi-token product units, particularly in compounding noun phrases.

**Table 10.** Examples in Named Entity Recognition for the GAM Task

| Sentence | Expected Entity | Predicted Output | Error Type |
|---|---|---|---|
| Specifically, the Trial Panel announced the division of the value of 15 assets, including: 5 apartments in the Avalon building (District 1)... | <address> Avalon building, District 1 </address> | <address> Avalon building </address> | Incomplete span / missing suffix |
| ...contributed capital in Binh Minh Co., Ltd. equivalent to over 3,700 m2 of land in Ba Ria - Vung Tau province... | <address> Ba Ria - Vung Tau province </address> | No label | Missed location |
| ...Sao Mai Port; the transferred amount... | <address> Sao Mai Port </address> | No label | Missed location |
| I'll take the size 2.16 m * 1.2 m... | <customer height> 2.16 m </customer height><product width> 1.2 m </product width> | <customer height> 2.16 m </customer height> | Partial numeric extraction |
| The feature of selling gold bars was transferred by this bank to be applied on the app... | <product unit> gold bars </product unit> | <product unit> bars </product unit> | Incomplete product span |
| Address: No. 226 Le Nin Street, Block 10, Quan Bau Ward, Vinh City, Nghe An. | <address> Address: No. 226 Le Nin Street, Block 10, Quan Bau Ward, Vinh City, Nghe An. </address> | <address> "No. 226 Le Nin Street </address> | Truncated address entity |

In task AI Fluent has encountered some errors (see table 11):

Nested Entity Handling: The model often fails to capture nested or overlapping entities. For instance, in a sentence containing both <allowance> and nested information like <ratio> or <money allowance>, the outer label may be correctly extracted but inner components are missed, or vice versa.

Label Inconsistency: Some well-defined categories (e.g., salary, transportation allowance) are missed altogether, despite strong lexical and syntactic cues. This suggests instability in label assignment when encountering compound financial expressions or domain-specific terminology.

Entity Type Confusion: There are instances where the model tags semantically similar phrases inconsistently. For example, “Phụ cấp ca đêm (night shift allowance) and “Phụ cấp hỗ trợ xe đi lại (transportation allowance) are structurally similar but receive inconsistent or missing tags, implying an inadequate representation of financial subtypes.

Underperformance on Coordinated Structures: The model struggles with multiple entities listed within the same sentence or clause, either predicting only one or failing to segment and recognize the structure correctly an issue exacerbated in financial texts which often list multiple items compactly.

**Table 11.** Examples in Named Entity Recognition for the AI Fluent Task

| Sentence | Expected Entity | Predicted Output | Error Type |
|---|---|---|---|
| The 1/4 overtime hours were approved but I see the overtime allowance amount added... Could you please check it for me... | <allowance> Overtime allowance </allowance><ratio> 1/4 </ratio> | <allowance> Overtime allowance </allowance> | Missing ratio |

| Still sent the wrong Tax ID | <Tax Identification Number> Tax ID </Tax Identification Number> | <Tax Identification Number> Tax ID </Tax Identification Number> | No Error |
|---|---|---|---|
| Which item is the difference between net income and total payment deducted from? – Missing night shift allowance 618,000 | <salary> Net income </salary><night shift allowance> Night shift allowance </night shift allowance><money allowance> 618,000 </money allowance> | <night shift allowance> Night shift allowance </night shift allowance><money allowance> 618,000 </money allowance> | Missing label salary |
| Could you check my salary, it is missing the transportation allowance. | <transportation allowance> Transportation allowance </transportation allowance> | No prediction | Missing label |

In addition to the general error categories, a manual review of 100 erroneous instances was conducted, identifying two primary types of errors: Partial Entity Match (accounting for 85% of cases, including span truncation and boundary errors) and Missing Entities (false negatives, representing 15%). This detailed analysis further clarifies the nature of the model's errors and strengthens understanding of the associated challenges, particularly in handling rare or complex entities. These findings have been incorporated into the manuscript to provide a clearer picture of the model's performance.

## 5. Conclusion

In this study, we proposed a high performance hybrid NER framework that effectively addresses the challenges of entity recognition in specialized Vietnamese domains with low-resource language. Our approach has combined rule-based techniques with deep learning methods and incorporates inference acceleration strategies such as Flash Attention, ONNX conversion, and KV Caching. This has enabled to achieve strong performance while maintaining efficiency in real-world deployments.

In addition, an optimized pipeline for industrial NER tasks, covering the full lifecycle from data collection and annotation to model selection and system design, our method achieved up to 90% F1-score on the Customer Service task, where entity boundaries are clear and align closely with colloquial Vietnamese. For more complex domains like GAM and AI Fluent, our RoBERTa-based models achieved F1-scores of 85% and 84%, respectively. Our system also showed significant improvement over a strong baseline on the rare Wildlife and PhoNER_Covid19 datasets. In terms of efficiency, KV Caching provided a 5.04x speedup, cutting the inference time for 9000-token sequences to just 5.02 seconds.

This study presents a high-performance hybrid Named Entity Recognition (NER) framework designed to address the inherent challenges of entity recognition within specialized Vietnamese domains characterized by low-resource constraints. By integrating rule-based techniques with deep learning methodologies and incorporating advanced inference acceleration strategies—including FlashAttention, ONNX conversion, and KV Caching—the framework achieves superior performance while maintaining operational efficiency for real-world deployments.

Furthermore, the proposed optimized pipeline encompasses the entire lifecycle of industrial NER tasks, ranging from data collection and annotation to model selection and system architecture. Experimental results demonstrate that the method attains an F1-score of up to 90% on the Customer Service task, where entity boundaries are distinct and align closely with colloquial Vietnamese patterns. In more complex domains, such as GAM and AI Fluent, the RoBERTa-based models achieve F1-scores of 85% and 84%, respectively. The system also exhibits significant performance gains over robust baselines on specialized datasets, including Wildlife and PhoNER_Covid19. Regarding computational efficiency, the implementation of KV Caching yields a 5.04x speedup, effectively reducing the inference time for 9,000-token sequences to 5.02 seconds.

## 6. Declarations

### 6.1. Author Contributions

D.M.D., Q.X.T., V.T.H., L.H.A., M.T.M.T., N.V.T., L.H.H., and V.N.V.; Methodology: M.T.M.T.; Software: D.M.D.; Validation: D.M.D., M.T.M.T., and V.N.V.; Formal Analysis: D.M.D., M.T.M.T., and V.N.V.; Investigation: D.M.D.; Resources: M.T.M.T.; Data Curation: M.T.M.T.; Writing Original Draft Preparation: D.M.D., M.T.M.T., and V.N.V.; Writing Review and Editing: M.T.M.T., D.M.D., and V.N.V.; Visualization: D.M.D.; All authors have read and agreed to the published version of the manuscript.

### 6.2. Data Availability Statement

The data presented in this study are available on request from the corresponding author.

### 6.3. Funding

The authors received no financial support for the research, authorship, and/or publication of this article.

### 6.4. Institutional Review Board Statement

Not applicable.

### 6.5. Informed Consent Statement

Not applicable.

### 6.6. Declaration of Competing Interest

The authors declare that they have no known competing financial interests or personal relationships that could have appeared to influence the work reported in this paper.